\documentstyle[12pt]{article}

\newcommand{\sptwo}{1.4}

\newcommand{\doublespace}{\edef\baselinestretch{\sptwo}\Large\normalsize}

\doublespace
\begin{document}
\begin{center}
{\bf Effective Linear Two-Body Method for Many-Body Problems
}\\
\renewcommand\thefootnote{\fnsymbol{footnote}}
{Yeong E. Kim \footnote{ e-mail: yekim$@$physics.purdue.edu,
fax: (765) 494 0706}
 and Alexander L. Zubarev\footnote{ e-mail: zubareva$@$physics.purdue.edu, fax: (765) 494 0706}}\\
Department of Physics, Purdue University\\
West Lafayette, Indiana  47907\\
{\bf Abstract}
\end{center}
\begin{quote}
 This paper reports a detailed description of the equivalent linear
 two-body method for the many body problem, which is based on an approximate 
reduction of the many-body Schr\"odinger equation by the use of a variational principle. 
 To test the accuracy of the method it has been applied to the one-dimensional $N$-body problem 
with pair-wise contact interactions (McGurie-Yang $N$-body problem)
and to the dilute Bose-Einstein condensation (BEC) of
atoms in harmonic traps at zero temperature.
For both cases, it is shown that the method gives excellent results for large $N$.
\end{quote}

\noindent
	
\pagebreak

{\bf I. Introduction}
\vspace{8pt}

In this paper we present an approximate method of obtaining the
eigenvalue solutions of the system of interacting $N$ bosons
using an equivalent two-body method similar to that used by Feshbach
and Rubinov [1] for the triton ($^3H$) three-body ($N$=3) bound state.
 They [1] used both the variational principle
and a reduced coordinate variable ( not the hyperradius)
to obtain an equivalent two-body equation for the three-body
bound state ($^3H$).
For many-body problems, use
of one 
reduced coordinate variable
 (the hyperradius [2])
was made to obtain equivalent two-body equations
by keeping only a finite sum of terms  of
the hyperspherical expansion with $K=K_{min}$
($K$ is the global angular momentum). This
method
has been applied to the ground state of the $N$-body 
system composed of distinguishable particles or of bosons
and also to nuclear bound states [3,4]. It was shown that the
method leads to the correctly behaved nuclear bound states in the
limit of large $A$ ($A$ is the nucleon number) [4]. 
Recently, it has been
used to describe the Bose-Einstein condensation (BEC) of atoms 
 in isotropic harmonic traps [5].
We note that Morse and Feshbah [6] have used hyperspherical coordinates for solving the problem of two interacting particles in a central field of force.
More references can be found in [7,8].

For the $N$-body system, our method for obtaining 
 the equivalent linear two-body (ELTB) equation consists of two
steps. The first step
is to give the $N$-body wave function $\Psi({\bf r_1},
{\bf r_2}, ...)$ a particular functional form
$$
\Psi({\bf r_1},{\bf r_2}, ...) \approx \tilde{\Psi}(\zeta_1,
\zeta_2, \zeta_3),
\eqno{(1})
$$
where $\zeta_1, \zeta_2, $and $\zeta_3$ are known functions. 
We limit $\zeta$'s to three variables in order to obtain the
ELTB equation.
 The second
step is to
derive an equation for $ \tilde{\Psi}(\zeta_1, \zeta_2,
\zeta_3)$  by requiring that $
\tilde{\Psi}$ must satisfy a variational principle
$$
\delta \int \tilde{\Psi}^\ast \tilde{\Psi}d \tau = 0
\eqno{(2)}
$$
with a subsidiary condition $\int \tilde{\Psi}^\ast \tilde{\Psi}d
\tau =1$.
This leads to a linear two-body equation from  which both
eigenvalues and eigenfunctions can be obtained.
The lowest
eigenvalue is an upper bound of the lowest eigenvalue of the
original N-body problem.

In Section II, we apply the method to the one-dimensional $N$-body
problem with pair-wise contact interactions (the McGuire-Yang $N$-body
problem [9, 10]) and demonstrate that our method is a very good approximation
for the case of large $N$. In Section III, we consider the Bose-Einstein
condensation (BEC) of atoms in isotropic harmonic traps. In Sections
IV and V we apply the method to the dilute BEC of atoms in anisotropic harmonic 
traps.
It was shown the method gives excellent results for large
$N$. In Section VI we investigate the stability of the BEC for the case of atoms with negative scattering length at zero temperature using our method. A summary 
and conclusions are given in Sec. VII.
\vspace{18pt}

{\bf II. McGuire-Yang N-body Problem}
\vspace{8pt}

There are only several known cases of exactly solvable three-body and
four-body problems. For the $N=3$ case it was shown [11] that
the Faddeev equations [12] for one-dimensional three-body
problem with pair-wise contact interactions are exactly solvable.
For the one-dimensional  $N=4$ case, analytical solutions 
of the four-body
Faddeev-Yakubovsky were obtained in [13].
We note that for nuclear three-body systems with short-range 
interactions, the  Schr\"odinger equation in three dimensions 
is reformulated into
the Faddeev equations [12] which have been solved numerically after
making
partial wave expansion [14] or without partial wave expansion
[15]. 
In the following, we consider an exactly solvable one-dimensional
$N$-body system as a test case for our method.

For the one-dimensional N-body problem with the Hamiltonian
$$
H=-\frac{1}{2}\sum_{i=1}^{N} d^2/dx_{i}^2+c \sum_{i<j} \delta(x_i-x_j),
$$
the Schr\"odinger equation
$$
H \Psi=E \Psi
$$
 is exactly solvable. The bound and
scattering states for this system have been found by McGuire [9]
and by Yang [10].

For the case $c<0$, there are bound states [9] for the system of
$N$ bosons with the wave function of the following form
$$
\Psi= \exp[(c/2) \sum_{i<j} \mid x_i-x_j \mid],
$$
 and the energy of this bound state is given by
$$ E=-c^2 N(N^2-1)/24.
\eqno{(3)}
$$

The McGuire-Yang (MY) N-body problem provides a unique possibility of
checking the validity and accuracy of various approximations made for the
Schr\"odinger equation describing $N$ particles interacting via short
range potential. In the following, we describe the
equivalent linear two-body equation (ELTBE) method 
 and compare its
solution with the exact solution, Eq. (3), of the MY problem.

For this case, we seek for eigenfunction $\Psi$ of $H$ in the form of
$$  
\Psi \approx \tilde{\Psi}( \rho), 
$$
where 
$$
 \rho= \sqrt{\frac{1}{N} \sum_{i<j} (x_i-x_j)^2}. 
$$
We now derive an ELTB equation for $ \tilde{\Psi}( \rho)$ by requiring that $
\tilde{\Psi}$ must satisfy a variational principle (2).
This requirement leads to the equation
$$ [- \frac{d^2}{d \rho^2}+\frac{(N-2)}{ \rho} \frac{d}{d \rho}+2V( \rho)]
\tilde{\Psi}( \rho) = 2E\tilde{\Psi}( \rho),
\eqno{(4)}
$$
where
$$
V( \rho) =\tilde{g}/ \rho,
\eqno{(5)}
$$
with
$$
 \tilde{g}= cN(N-1) \frac{ \Gamma (N/2-1/2)}{2 \sqrt{2 \pi } \Gamma
(N/2-1)}
\eqno{(6)}
$$
(see Appendix A for details).
 Eq. (4) with the Coulomb like potential $V( \rho)= \tilde{g}/ \rho$,
Eq. (5), can be solved analytically.

Instead of the parameter E and variable $\rho$ in the Schr\"odinger equation (4),
we introduce the new quantities
$$
z= \frac{2 \tilde{g}}{ \eta} \rho,\; \; 
\eta=\tilde{g} \sqrt{ \frac{1}{-2E}}
\eqno{(7)}
$$
For negative $\tilde{g}$ and negative energies, $\eta$ is real negative number.
On making the substitutions (7), Eq. (4) becomes
$$
\frac{d^2 \tilde{\Psi}}{dz^2}+\frac{(N-2)}{z} \frac{d \tilde{\Psi}}{dz}+[-\frac{1}{4}-\frac{\eta}{z}]\Psi=0.
\eqno{(8)}
$$
To calculate the asymptotic behavior of $ \tilde{\Psi}$ for large $z$, we omit 
from Eq. (8) the terms in $1/z$, and $1/z^2$, and obtain the equation
$$
\frac{d^2 \tilde{\Psi}}{dz^2}=\frac{1}{4} \tilde{\Psi},
$$
 which shows that $\tilde{\Psi}$ behaves as $exp[-z/2]$, a correct asymptotic 
form.

After substitution 
$$
\tilde{\Psi}(z)=e^{-z/2}y(z)
$$
Eq. (8) becomes
$$ z \frac{d^2y}{dz^2}+(N-2-z)\frac{dy}{dz}-( \eta+\frac{N-2}{2})y=0.
\eqno{(9)}
$$
The solution of this equation (finite for $z=0$) is the confluent 
hypergeometric function
$$
y=F(\eta+\frac{N-2}{2},N-2,z).
\eqno{(10)}
$$
A solution which satisfies the condition at infinity is obtained only for negative integer (or zero) values of $\eta+\frac{N-2}{2}$,
$$
\eta+\frac{N-2}{2}=-n, (n=0,1,...).
\eqno{(11)}
$$
 From the definition, Eq.(7), of the parameter $\eta$, we find
$$
E_{n}(N)=-2\frac{ \tilde{g}^2}{(N-2+2n)^2}.
\eqno{(12)}
$$
 Using Eq.(6), we have the following expression
for ground state energy 
$$
E_{0}(N)=-\frac{c^2}{4 \pi}[ \frac{N(N-1)
\Gamma(N/2-1/2)}{(N-2) \Gamma(N/2-1)}]^2.
\eqno{(13)}
$$
We note that our system has only one bound state, $n=0,$ consistent
with the exact solution, Eq.(3). States with $n \neq 0$
are not bound states because $ \mid E_{n \neq 0}(N) \mid < \mid E_{0}(N-1) \mid
$.

In the case of large $N$, using the asymptotic formulas for
$\Gamma$ function,
$$
\lim_{\mid z\mid \rightarrow \infty} \frac{\Gamma(z+\alpha)}
{\Gamma(z+\beta)}=z^{\alpha-\beta}(1+O(\frac{1}{z})),
$$
we obtain 
$$E_{0}(N)=-\frac{c^2}{8 \pi}N^3
$$ 
for the leading term of Eq. (13).
 On the other hand we have for the large $N$ case from Eq. (3), 
$$E_=
-\frac{c^2}{24}N^3(1+O(\frac{1}{N^2})).$$
Therefore, for the McGuire-Yang N-body problem,
we have demonstrated that the 
ELTBE method, Eqs. (4 - 6), is a very good
approximation for the case of large $N$
(the relative error for binding energy is about 4.5 \%).
Furthermore, our approximation, Eq. (13), agrees remarkably well
with the exact value, Eq. (3), for any $N$ (the maximum value of relative 
error for the binding
energy occurs for $N=3$ and is about 10 \%).

\vspace{8pt}

{\bf III. Isotropic Trap}
\vspace{8pt}

 In this section, we consider $N$ identical bosonic atoms confined in a
harmonic isotropic trap with the following Hamiltonian

$$
H=-\frac{\hbar^2}{2m}  \sum_{i=1}^{N} \Delta_{i}+\frac{1}{2}
m \omega^2 \sum_{i=1}^{N}r_i^2
+\sum_{i<j}V_{int}({\bf r}_i-{\bf r}_j).
\eqno{(14)}
$$

For the eigenfunction $\Psi$ of $H$, we assume that the solution for
$\Psi$ has the following form

$$
\Psi(\vec{r}_1, ...\vec{r}_N) \approx \frac{ \tilde{ \Psi}( \rho)}{ \rho^{(3N-1)/2}},
\eqno{(15)}
$$
where
$$
\rho=\sum_{i=1}^{N}r_i^2.
\eqno{(16)}
$$
We now derive an ELTB equation for $\tilde{\Psi}$
 by requiring
that $\tilde{ \Psi}$ must satisfy the variational principle (2).
This requirement leads to the equation
$$
\tilde{H} \tilde{\Psi}=E \tilde{\Psi},
\eqno{(17)}
$$
where
$$
\tilde{H}=-\frac{\hbar^2}{2m} \frac{d^2}{d \rho^2}+\frac{m}{2} \omega^2 \rho^2
+\frac{\hbar^2}{2m}\frac{(3N-1)(3N-3)}{4 \rho^2}+V(\rho),
\eqno{(18)}
$$
with
$$
V(\rho)=\frac{N(N-1)}{\sqrt{2 \pi}} \frac{\Gamma(3N/2)}{\Gamma(3N/2-3/2)}
\frac{1}{\rho^3} \int_{0}^{\sqrt{2} \rho}V_{int}(r) (1-\frac{r^2}{2 \rho^2})^
{(3N/2-5/2)}r^2 dr
\eqno{(19)}
$$
(see Appendix B for details).

We note that Eq. (17) is exactly the form of the
Schr\"odinger two-body equation in which 
 a centrifugal potential energy is given by $
(N-1)(N-3)/(4 \rho^2)$ with identification of angular momentum
quantum number
$l=(N-1)/2$.

In the dilute condensate case
$$
\rho \gg r_A,
\eqno{(20)}
$$
where $r_A$ is an atom-atom interaction range. Hence we can use for
$V(\rho)$, Eq.(19),
 the following approximation
$$
V(\rho)=\frac{N(N-1)}{\sqrt{2 \pi}} \frac{\Gamma(3N/2)}{\Gamma(3N/2-3/2)}
\frac{1}{\rho^3} \int_{0}^{\infty}V_{int}(r)r^2 dr,
\eqno{(21)}
$$
which is
proportional to the scattering length $a_B$ in the Born approximation for binary 
collisions 
$$
a_B=\frac{m}{4 \pi \hbar^2} \int_{0}^{\infty}V_{int}(r)r^2 dr.
\eqno{(22)}
$$
Since the actual atom-atom interaction in condensate is much larger than the
scattering
energy, it is not possible to use perturbation theory to describe the scattering. However, since large changes in the wave function only occur over very small 
distances and since the wave function outside the range of interaction is only
slightly changed by the interaction, Fermi [16] realized that it is possible to 
introduce a pseudopotential, which can be used to calculate small changes in the wave function outside the range of interaction by perturbation theory.

Using the above argument, let us introduce a pseudopotential $U$, so that the 
two-body Schr\"odinger 
equation becomes
$$
-\frac{\hbar^2}{m} \bigtriangleup \psi+U \psi=E \Psi,
\eqno{(23)}
$$
with
$$
U = \left\{ \begin{array}{ll}
                U_0 & \mbox{if $r< \eta$,}\\
                 0  & \mbox{otherwise,}
                \end{array}
                 \right. 
\eqno{(24)}
$$
where $\eta$ is some distance chosen so that $\eta \gg \mid a \mid$, 
$\eta \gg r_A$, and
$\eta \ll \tilde{\rho}$, with the scattering length $a$ and 
range of condensate
$\tilde{\rho}$.

We hope to find $U$ so that the exact atom-atom wave function for $r > \eta$
will be given by the solution of Eq. (23) with amplitude $f(\theta)$
given by the Born approximation applied to $U$

$$
f_B(\theta)=-\frac{m}{4 \pi \hbar^2}\int d^3rU(r)e^{i(\vec{k}_i-\vec{k}_f)
\cdot \vec{r}}.
\eqno{(25)}
$$
Since
we know that the exact $f(\theta)$ is independent of 
$\theta$ for our very low energy case, the range $\eta$  of the pseudopotential 
$U(r)$ must satisfy

$$
k \eta \ll 1,
\eqno{(26)}
$$
where $k=\sqrt{mE/ \hbar^2}$. In this case Eq. (25) reduces to
$$
f_B(\theta)=-a_B=-\frac{m}{3\hbar^2}U_0 \eta^3.
\eqno{(27)}
$$
In order for the Born approximation to be valid we require the following 
relation
$$
\frac{a-a_B}{a} \ll 1,
\eqno{(28)}
$$
where $a$ is the exact scattering length with pseudopotential $U(r)$.

Combining Eqs. (26-28) we find
$$
\eta_0 \ll \eta \ll \frac{1}{k},
\eqno{(29)}
$$
where $\eta_0$ is fixed from Eq.(28).

 Therefore, the
 concept of pseudopotential may be used
as long as the energy of relative motion, E, is small
$$
E \ll \frac{\hbar^2}{m \eta_0^2}
\eqno{(30)}
$$
where $\eta_0$ is fixed from the condition $\frac{a-a_B}{a} \leq 0.1$.
For the $^7Li$ case we have 
$$
\eta_0=0.2\times10^{-5} cm,
\eqno{(31)}
$$
and for the case of $^{87}Rb$ atoms we have
$$
\eta_0=0.7\times10^{-5}cm.
\eqno{(32)}
$$
Eqs.(30-32) imply that the temperature of the condensate must be 
considerably lower than $10^{-4} K$ for the $^7Li$ condensate and 
$7.6 \times 10^{-7} K$ for the $^{87}Rb$ condensate.
\vspace{18pt}

\pagebreak

{\bf IV. Anisotropic Trap}
\vspace{8pt}

In this section,
we consider $N$ identical bosonic atoms confined in a
harmonic anisotropic trap with the following Hamiltonian

$$
H=-\frac{\hbar^2}{2m} \sum_{i=1}^{N} \Delta_{i}+\frac{1}{2}
\sum_{i=1}^{N}m(\omega_x^2x_i^2+\omega_y^2y_i^2+\omega_z^2z_i^2)
+\sum_{i<j}V_{int}(\vec{r}_i-\vec{r}_j),
\eqno{(33)}
$$
For eigenfunction $\Psi$ of $H$, we assume the solution for
$\Psi$ has the following form
$$
\Psi(\vec{r}_1, ...\vec{r}_N) \approx \frac{ \tilde{\Psi}(x,y,z)}{
(xyz)^{(N-1)/2}},
\eqno{(34)}
$$
where
$$
x^2=\sum_{i=1}^{N}x_i^2,\  y^2=\sum_{i=1}^{N}y_i^2,\
z^2=\sum_{i=1}^{N}z_i^2.
\eqno{(35)}
$$

We now derive an equation for $\tilde{\Psi}(x,y,z)$ by requiring
that $\tilde{\Psi}(x,y,z)$ must satisfy the variational principle
(2). This requirement leads to the equation
$$
\tilde{H} \tilde{\Psi}=E\tilde{\Psi},
\eqno{(36)}
$$
where
$$
\begin{array}{rcl}
&~& \tilde{H}=-\frac{\hbar^2}{2m}(\frac{\partial^2}{\partial x^2}+\frac{\partial^2}
{\partial y^2}+\frac{\partial^2}{\partial z^2})+\frac{m}{2}(\omega_x^2x^2
+\omega_y^2y^2+\omega_z^2z^2)\\ 
&~& \\
&~& +\frac{\hbar^2}{2m}\frac{(N-1)(N-3)}{4}(
\frac{1}{x^2}+\frac{1}{y^2}+\frac{1}{z^2})+V(x,y,z),
\end{array}
\eqno{(37)}
$$
with 
$$
V(x,y,z)=\frac{N(N-1)}{2(2 \pi)^{3/2}}(\frac{\Gamma(N/2)}{\Gamma((N-1)/2)})^{3}
\frac{1}{xyz}G(x,y,z),
\eqno{(38)}
$$
and
$$
\everymath={\displaystyle}
\begin{array}{rcl}
&~& G(x,y,z)=\int_{-\sqrt{2}x}^{\sqrt{2}x}dx^{\prime}\int_{-\sqrt{2}y}^{\sqrt{2}y}dy^{\prime}
\int_{-\sqrt{2}z}^{\sqrt{2}z}dz^{\prime}V_{int}(\sqrt{(x^{\prime})^2+(y^{\prime})^2+(z^{\prime})^2}) \\
&~& \\
&~& \times ((1-\frac{(x^{\prime})^2}{2x^2})(1-\frac{(y^{\prime})^2}{2y^2})(1-
\frac{(z^{\prime})^2}{2z^2}))^{\frac{N-3}{2}}
\end{array}
\eqno{(39)}
$$
(see Appendix C for details).
To the best of our knowledge,
Eqs. (38) and (39) have not been discussed in the literature.

In the dilute condensate case
$$
x \gg r_A,\ \ y \gg r_A,\ \ z \gg r_A
\eqno{(40)}
$$
we can use for V(x,y,z) the following approximation
$$
\everymath={\displaystyle}
\begin{array}{rcl}
&~& V(x,y,z)=\frac{N(N-1)}{2(2 \pi)^{3/2}}(\frac{\Gamma(N/2)}{\Gamma((N-1)/2)})^{3}
\frac{1}{xyz} \\
&~& \\
&~& \times \int_{-\sqrt{2}x}^{\sqrt{2}x}dx^{\prime}\int_{-\sqrt{2}y}^{\sqrt{2}y}dy^{\prime}
\int_{-\sqrt{2}z}^{\sqrt{2}z}dz^{\prime}V_{int}(\sqrt{(x^{\prime})^2+(y^{\prime})^2+(z^{\prime})^2}).
\end{array}
\eqno{(41)}
$$
 Since we have the following relation for the case of large $x$, $y$, and $z$, 
$$
\int_{-\sqrt{2}x}^{\sqrt{2}x}dx^{\prime}\int_{-\sqrt{2}y}^{\sqrt{2}y}dy^{\prime}
\int_{-\sqrt{2}z}^{\sqrt{2}z}dz^{\prime}V_{int}(\sqrt{(x^{\prime})^2+(y^{\prime})^2+(z^{\prime})^2})=
4 \pi\int_{0}^{\infty}V_{int}(r)r^2 dr
\eqno{(42)}
$$
V(x,y,z) is proportional to the scattering length in the 
Born approximation for binary collisions
$$
V(x,y,z)=\frac{g}{xyz},
\eqno{(43)}
$$
with
$$
g=\frac{a_B\hbar^2N(N-1)}{\sqrt{2\pi}m}(\frac{\Gamma(N/2)}{\Gamma((N-1)/2)})^{3}.
\eqno{(44)}
$$
 Therefore, it is reasonable to replace expression (44) with a corresponding 
expression proportional to the exact scattering length $a$ for binary collisions
(Landau replacement [17]).
$$
g=\frac{a\hbar^2N(N-1)}{\sqrt{2\pi}m}(\frac{\Gamma(N/2)}{\Gamma((N-1)/2)})^{3}.
\eqno{(45)}
$$
This approximation is equivalent to the following approximation for $V_{int}$
$$
V_{int}(\vec{r}_i-\vec{r}_j)=\frac{4 \pi \hbar^2a}{m}\delta(\vec{r}_i-\vec{r}_j),
\eqno{(46)}
$$
 which is the Fermi pseudopotential [16].

 For the positive scattering length case, $a>0$, we
look for the solution of Eq. (36) of the form
$$
\tilde{\Psi}(x,y,z)=\sum_{i,j,k}c_{ijk} \Phi_i^{(1)}(x)
\Phi_j^{(2)}(y) \Phi_k^{(3)}(z),
\eqno{(47)}
$$
 where 
$c_{ijk}$ are solutions of the following equations
$$
\sum_{l,m,n}H_{ijk,lmn}c_{lmn}=E \sum_{l,m,n} \lambda_{ijk,lmn}c_{lmn}
\eqno{(48)}
$$
with
$$
H_{ijk,lmn}=<\Phi_i^{(1)}\Phi_j^{(2)}\Phi_k^{(3)} \mid \tilde{H} \mid
\Phi_l^{(1)}\Phi_m^{(2)}\Phi_n^{(3)}>,
\eqno{(49)}
$$
and
$$
\lambda_{ijk,lmn}=<\Phi_i^{(1)}\Phi_j^{(2)}\Phi_k^{(3)} \mid
\Phi_l^{(1)}\Phi_m^{(2)}\Phi_n^{(3)}>.
\eqno{(50)}
$$
Using
$$
\everymath={\displaystyle}
\begin{array}{rcl}
\Phi_i^{(1)}(x)=x^{(N-1)/2} \exp[-m \tilde{\omega}(x/ \alpha_i)^2/(2\hbar)],\\
\Phi_j^{(2)}(y)=y^{(N-1)/2} \exp[-m \tilde{\omega}(y / \beta_j)^2/(2\hbar)],
\end{array}
$$
and
$$
\everymath={\displaystyle}
\begin{array}{rcl}
\Phi_k^{(3)}(z)=z^{(N-1)/2} \exp[-m \tilde{\omega}(z/ \gamma_k)^2/(2\hbar)],
\end{array}
$$
we have
$$
\everymath={\displaystyle}
\begin{array}{rcl}
&~& H_{ijk,lmn}=\frac{\hbar
\tilde{\omega}N\lambda_{ijk,lmn}}{2}[\frac{1+\alpha_i^2
\alpha_l^2 \alpha_x^2}{\alpha_i^2+\alpha_l^2}+ \frac{1+\beta_j^2
\beta_m^2 \alpha_y^2}{\beta_j^2+\beta_m^2}+ \frac{1+\gamma_k^2
\gamma_n^2 \alpha_z^2}{\gamma_k^2+\gamma_n^2} \\
&~& \\
&~& + \tilde{g}\frac{ \sqrt{(\alpha_i^2+\alpha_l^2)
(\beta_j^2+\beta_m^2)(\gamma_k^2+\gamma_n^2)}}{\alpha_i \alpha_l \beta_j \beta_m
 \gamma_k \gamma_n}],
\end{array}
\eqno{(51)}
$$
and
$$
\lambda_{ijk,lmn}=[\frac{8 \alpha_i \alpha_l \beta_j \beta_m
\gamma_k \gamma_n}{(\alpha_i^2+\alpha_l^2)(\beta_j^2+\beta_m^2)
(\gamma_k^2+\gamma_n^2)}]^{N/2}.
\eqno{(52)}
$$
 with $\tilde{g}=\frac{(N-1)}{2 \sqrt{2}N}\tilde{n}$, $\tilde{n}=
2\sqrt{\tilde{\omega}m/(2
\pi \hbar)}Na$, $\tilde{\omega}=(\omega_x \omega_y \omega_z)^{\frac{1}{3}}
$, $\alpha_x=\omega_x/\tilde{\omega}$,
$\alpha_y=\omega_y/\tilde{\omega}$, and
$\alpha_z=\omega_z/\tilde{\omega}$.

For the case of large $N$, $\lambda_{ijk,lmn}$ reduces to the $\delta$-function
$$
\lambda_{ijk,lmn}\approx \delta_{il} \delta_{jm} \delta_{kn},
\eqno{(53)}
$$
and hence
$$
H_{ijk,lmn} \approx E \delta_{il} \delta_{jm} \delta_{kn}.
\eqno{(54)}
$$
Using Eq. (54) we have for the ground state energy
$$
E=\frac{\hbar \tilde{\omega} N}{2}[\frac{1}{2 \alpha^2}+\frac{1}{2 \beta^2}
+\frac{1}{2 \gamma^2}+\frac{\alpha_x^2}{2}\alpha^2+\frac{\alpha_y^2}{2}\beta^2
+\frac{\alpha_z^2}{2}\gamma^2+\tilde{g} \frac{2 \sqrt{2}}{\alpha \beta \gamma}],
\eqno{(55)}
$$
where parameters $\alpha, \beta,$ and $\gamma$ are solutions of the following 
equations
$$
\frac{\partial E}{\partial \alpha}=\frac{\partial E}{\partial \beta}
=\frac{\partial E}{\partial \gamma}=0.
\eqno{(56)}
$$
 For the case of large $N$ we can neglect the kinetic energy term in Eq. (55)
$$
E=\frac{\hbar \tilde{\omega} N}{2}[\frac{\alpha_x^2}{2}\alpha^2+
\frac{\alpha_y^2}{2}\beta^2+\frac{\alpha_z^2}{2}\gamma^2+\tilde{g} \frac{2 
\sqrt{2}}{\alpha \beta \gamma}]
\eqno{(57)}
$$
Substitution Eq. (57) into Eq. (56) gives
$$
\alpha_x^2\alpha^2=\frac{2 \sqrt{2}\tilde{g}}{\alpha \beta \gamma},\; \;
\alpha_y^2\beta^2=\frac{2 \sqrt{2}\tilde{g}}{\alpha \beta \gamma},\; \;
\alpha_z^2\gamma=\frac{2 \sqrt{2}\tilde{g}}{\alpha \beta \gamma}.
\eqno{(58)}
$$
Solutions of these equations
$$
\alpha=\frac{(2 \sqrt{2}\tilde{g}\alpha_x\alpha_y\alpha_z)^{1/5}}{\alpha_x},\; \;
\beta=\frac{(2 \sqrt{2}\tilde{g}\alpha_x\alpha_y\alpha_z)^{1/5}}{\alpha_y},\; \;
\gamma=\frac{(2 \sqrt{2}\tilde{g}\alpha_x\alpha_y\alpha_z)^{1/5}}{\alpha_z}
$$
give for the ground state energy
$$
\frac{E}{N \hbar \tilde{\omega}}=\frac{5}{4}\tilde{n}^{\frac{2}{5}}
\eqno{(59)}
$$
We note that Eq.(59) is the exact ground state solution of Eq.(36)
for large $N$. For the case of large $N$ we can obtain an
essentially exact expression for the ground state energy
by neglecting the kinetic energy term in the
Ginzburg-Pitaevskii-Gross (GPG) equation [19]
 (the Thomas-Fermi
approximation [18]) as
$$
\frac{E_{TF}}{N \hbar \tilde{\omega}}=\frac{5}{7}(\frac{15}{8}
\sqrt{ \pi})^{\frac{2}{5}} \tilde{n}^{\frac{2}{5}}
\eqno{(60)}
$$

Comparing Eq. (59) with Eq. (60), we can see that for the case of  
large $N$, the
ELTBE method is a very good approximation, 
with a relative 
error of about 8$\%$
for the binding energy. 
\vspace{18pt}

{\bf V. Large N Limit.}
\vspace{8pt}

After we have obtained Eq. (37) the next step is to make a proper choice for 
the 
large N limit of the Hamiltonian. To do this let us rescale variables x, y, z
$$
x=N^{1/2}\tilde{x},\; y=N^{1/2}\tilde{y},\; z=N^{1/2}\tilde{z}.
\eqno{(61)}
$$
We can rewrite Eq. (36) as
$$
\begin{array}{rcl}
&~& [-\frac{\hbar^2}{2mN^2}(\frac{\partial^2}{\partial \tilde{x}^2} +\frac{\partial
^2} {\partial  \tilde{y}^2} +\frac{\partial^2}{\partial \tilde{z}^2})
+\frac{m}{2}(\omega_x^2 \tilde{x}^2+\omega_y^2\tilde{y}^2+\omega_z^2
\tilde{z}^2)\\
&~& \\
&~& +\frac{\hbar^2}{2m}\frac{(N-1)(N-3)}{4N^2}(
\frac{1}{\tilde{x}^2}+\frac{1}{\tilde{y}^2}+\frac{1}{\tilde{z}^2})+
\frac{a \hbar^2 N(N-1)}{mN\sqrt{2\pi}}
(\frac{\Gamma(N/2)}{\Gamma((N-1)/2)N^{1/2}})^3\frac{1}{\tilde{x}\tilde{y}
\tilde{z}}]\tilde{\Psi}=\frac{E}{N} \tilde{\Psi}.
\end{array}
\eqno{(62)}
$$
In the large N limit, $\frac{(N-1)(N-3)}{N^2}$ is of the order of unity and 
$(\frac{\Gamma(N/2)}{\Gamma((N-1)/2)N^{1/2}})^3$ is of the order of $(1/2)^{3/2}$,
 and Eq. (62) simplifies to
$$
[-\frac{\hbar^2}{2mN^2}(\frac{\partial^2}{\partial \tilde{x}^2} +
\frac{\partial^2}
{\partial  \tilde{y}^2} +\frac{\partial^2}{\partial \tilde{z}^2})
+V_{eff}(\tilde{x},\tilde{y},\tilde{z})]\tilde{\Psi}=\frac{E}{N} \tilde{\Psi},
\eqno{(63)}
$$
where
$$
V_{eff}(\tilde{x},\tilde{y},\tilde{z})=\frac{m}{2}(\omega_x^2 \tilde{x}^2+
\omega_y^2\tilde{y}^2+\omega_z^2\tilde{z}^2)+\frac{\hbar^2}{8m}(
\frac{1}{\tilde{x}^2}+\frac{1}{\tilde{y}^2}+\frac{1}{\tilde{z}^2})+
\frac{aN\hbar^2}{4m\sqrt{\pi}\tilde{x}\tilde{y}\tilde{z}}.
\eqno{(64)}
$$
Equation (63) describes the motion of a particle with an effective mass $mN^2$
 in an effective potential $V_{eff}(\tilde{x},\tilde{y},\tilde{z})$.
Therefore when $N \rightarrow \infty$, the effective mass of the particle becomes infinitely large and then the particle may be assumed to remain essentially stationary at the absolute minimum of $V_{eff}(\tilde{x},\tilde{y},\tilde{z})$.
Quantum fluctuations are unimportant in this limit and the most significant contribution to the ground state energy is given by
$$
E=NV_{eff}(x_0,y_0,z_0),
\eqno{(65)}
$$
where $x_0,y_0,z_0$ are to be obtained from
$$
\frac{\partial V_{eff}(x_0,y_0,z_0)}{\partial x_0}=
\frac{\partial V_{eff}(x_0,y_0,z_0)}{\partial y_0}=
\frac{\partial V_{eff}(x_0,y_0,z_0)}{\partial z_0}=0.
\eqno{(66)}
$$
Obviously Eq. (65) fails if the effective potential does not possess a minimum.

Instead of variables $\tilde{x}, \tilde{y}, \tilde{z}$ we introduce the new quantities
$$
x_t=\sqrt{\frac{m\tilde{\omega}}{\hbar}}\tilde{x},\;
y_t=\sqrt{\frac{m\tilde{\omega}}{\hbar}}\tilde{y},\;
z_t=\sqrt{\frac{m\tilde{\omega}}{\hbar}}\tilde{z}.
\eqno{(67)}
$$
On making the substitutions (67),Eqs. (64) and (66) become
$$
V_{eff}(x_t,y_t,z_t)=\frac{\hbar \tilde{\omega}}{2}[(\alpha_x^2x_t^2+
\alpha_y^2y_t^2+\alpha_z^2z_t^2)+\frac{1}{4}(\frac{1}{x_t^2}+\frac{1}{y_t^2}+
\frac{1}{z_t^2})+\frac{\tilde{n}}{2^{3/2}}\frac{1}{x_t y_t z_t}],
\eqno{(68)}
$$
with
$$
\everymath={\displaystyle}
\begin{array}{rcl}
2\alpha_x^2x_t+\frac{1}{2x_t^3}=\frac{\tilde{n}}{x_t^2y_tz_t},\\
2\alpha_y^2y_t+\frac{1}{2y_t^3}=\frac{\tilde{n}}{y_t^2x_tz_t},
\end{array}
\eqno{(69)}
$$
and
$$
\everymath={\displaystyle}
\begin{array}{rcl}
2\alpha_z^2z_t+\frac{1}{2z_t^3}=\frac{\tilde{n}}{z_t^2x_ty_t}.
\end{array}
$$
In the case of large $\tilde{n}=2\sqrt{\tilde{\omega}m/2\pi\hbar}Na$
we can neglect $\frac{1}{4}(\frac{1}{x_t^2}+\frac{1}{y_t^2}+
\frac{1}{z_t^2})$. In this case, solutions of Eq. (69)
$$
x_t^2=\frac{\tilde{n}^{2/5}}{2\alpha_x^2},\; \;
y_t^2=\frac{\tilde{n}^{2/5}}{2\alpha_y^2},\; \;
z_t^2=\frac{\tilde{n}^{2/5}}{2\alpha_z^2},
\eqno{(70)}
$$
give for the ground state energy, Eq. (65) 
$$
E/(N\hbar \omega)= \frac{5}{4} \tilde{n}^{2/5},
$$
which is identical to (59). Hence we show the semiclassical nature of the 
large $N$ approximation (59). Corrections to the result of the large $N$ limit
for the finite $N$ case
may be obtained by incorporating in the theory the quantum fluctuations around 
the classical minimum [20].
\vspace{18pt}

{\bf VI. Stability of BEC}
\vspace{8pt}

When the scattering length is negative, the effective
interaction between atoms is attractive. It has been claimed that
the BEC in free space is impossible [21] because the attraction
makes the system tend to an ever dense phase. For $^7Li$, the
$s$-wave scattering length is $a=(-14.5 \pm 0.4) \AA$ [22]. For
bosons trapped in an external potential there may exist a
metastable BEC state with a number of atoms below the critical
value $N_{cr}$ [23-31].  

For the $a<0$ case, we can see that potential energy  in Eq. (37),
$$
W(x,y,z)=\frac{m}{2}(\omega_x^2x^2+\omega_y^2y^2+\omega_z^2z^2)+
\frac{\hbar^2}{2m}\frac{(N-1)(N-3)}{4}(\frac{1}{x^2}+\frac{1}{y^2}+
\frac{1}{z^2})-
\frac{ \mid g \mid }{xyz},
\eqno{(71)}
$$
for $N<N_{cr}$ has a single metastable minimum which leads to the
metastable BEC state. We note that for the case of large $N_{cr}$,
the ELTBE method leads to the same $N_{cr}$ as the variational GPG 
stationary
theory [30]. To show this, let us consider an anisotropic trap, $
\omega_x=\omega_y=\omega_\bot$, $\omega_z=\lambda \omega_\bot $.
Local minimum conditions $\hat{A} > 0$, where $\hat{A}$ is a
matrix with matrix elements $A_{ij}=\partial^2W/ \partial x_i
\partial x_j$, can be written for this case as
$$
n^2/2 \delta_\bot^2 \delta_z^{\frac{1}{2}}-n-
\lambda ^2 \delta_\bot \delta_z^{\frac{3}{2}}/32+O(\frac{1}{N})
< 0,
\eqno{(72)}
$$
where 
$
\delta_z=(2m \omega_\bot/\hbar N_{cr})z^2,
\delta_\bot=(2m \omega_\bot/ \hbar N_{cr})x^2,$ and
$$
n=2(m
\omega_\bot/2 \pi \hbar)^{1/2}N_{cr}\mid a \mid.
\eqno{(73)}
$$
Setting the left-hand side of Eq. (72) to zero and
neglecting $O(\frac{1}{N})$ terms, we obtain the following
equations for $N_{cr}$ 
$$ 
\everymath={\displaystyle}
\begin{array}{rcl}
1 - 2\delta_\bot^2 = \delta_\bot^2 (1 + 8
\frac{\delta_z}{\delta_\bot} \lambda^2)^{1/2}, \\
1 - \lambda^2 \delta_z^2 = \delta_z\delta_\bot[1 + (1 + 8
\frac{\delta_z}{\delta_\bot} \lambda^2)^{1/2}],
\end{array}
\eqno{(74)}
$$
and
$$
\everymath={\displaystyle}
\begin{array}{rcl}
n = \delta_z^{1/2}\delta_\bot^2[1 + (1 + 8
\frac{\delta_z}{\delta_\bot} \lambda^2)^{1/2}] .
\end{array}
$$

\noindent
Eqs. (74) are exactly the same as equations for determining $N_{cr}$
obtained from the variational GPG approach [30].
In this reference [30], it was found that the numerical solution of Eqs. (57) for 
$0 \leq \lambda \leq 1$  can be interpolated as 
$$ n = e^{-(\alpha + \beta \lambda^2)}
\eqno{(75)}
$$
with $\alpha$ = 0.490419, $\beta$ = 0.149175.
Using Eqs. (74) and (75) we have
$$
N_{cr}= (\frac{2 \pi \hbar}{\omega_\bot^{(o)}m})^{1/2}~~
\frac{e^{-(\alpha + \beta \lambda^2)}}{2|{\it a}|}
\eqno{(76)}
$$
For an isotropic trap ($\lambda$ = 1) we obtain from Eqs. (57)
$\delta_z = \delta_\bot = 5^{-1/2}$ and $n=5^{-1/4}0.8 \approx
0.535$ which are in agreement with the results of Refs. [24, 25].
Taking the experimental values of $^7Li$ trap parameters [32], $\omega_\bot /2
\pi$ = 152 Hz, and $\omega_z/2 \pi$ = 132 Hz we obtain $
N_{cr}$= 1456. This value of $ N_{cr}$ is consistent with
theoretical predictions [27-31] and is in agreement with those
observed in a recent experiment [32].

We note that the ELTBE method for a
general anisotropic trap can be improved using a generalization
of the hyperspherical expansion
$$
\Psi({\bf r}_1, ...{\bf
r}_N)=\sum_{\stackrel{K_x,K_y,K_z,}{ \nu_x,\nu_y, \nu_z}}
\Psi_{K_x,K_y,K_z}^{  \nu_x,\nu_y, \nu_z}(x,y,z)Y^{\nu_x}_{K^{\nu_x}_x}(\Omega_x)Y^{\nu_y}_{K_y}(\Omega_y)Y^{\nu_z}_{K_z}(\Omega_z),
\eqno{(77)}
$$
where the hyperspherical harmonics $Y^{\nu_x} _{K_x}(\Omega_x)$, $
Y^{\nu_y}_{K_y}(\Omega_y)$, and $Y^{\nu_z}_{K_z}(\Omega_z)$ 
are eigenfunctions of the
angular parts of the Laplace operators $\sum_{i=1}^{N} \frac{\partial^2}
{\partial x_i^2}$, $\sum_{i=1}^{N} \frac{\partial^2}{\partial
y_i^2}$, and $\sum_{i=1}^{N} \frac{\partial^2}{\partial z_i^2}$,
respectively. However, we do not expect a fast convergence of the
expansion Eq. (77) because of nonuniformity  
of the convergence of the expansion of
$
\sum_{i<j}V_{int}(\vec{r}_i-\vec{r}_j)$ in $x$, $y$, and $z$.
\vspace{18pt}

{
\bf VII. Summary and conclusions}
 \vspace{8pt}

In summary, we have presented a method for obtaining an 
equivalent linear two-body equation from the Schr\"odinger equation
 for the system of $N$ bosons, using reduced variables and
variational principle. To access the accuracy of the method it has 
  been applied
to the McGuire-Yang N-body problem for which the exact solutions are known.
Our method gives excellent results compared with exact solutions.
The method has been applied also to the dilute Bose-Einstein condensation
in anisotropic harmonic traps at zero
temperature for both positive and negative scattering length. For large $N$,
our method gives excellent results for all these cases.

 \pagebreak

{\bf Appendix A}

\vspace{8pt}

In this Appendix, we derive Eqs. (5)-(6).
To calculate $V(\rho)$ for McGuire-Yang problem we start from definition
$$
V(\rho)=\frac{N(N-1)}{2}\int_{- \infty}^{\infty}dx_1...\int_{- \infty}^{\infty}dx_{N-1}\int_{- \infty}^{\infty}dtV_{int}(\sqrt{2}x_1)e^{i\sum_{n=1}^{N-1} (x_n^2-\rho^2)t}/ \Omega,
\eqno{(A.1)}
$$
where 
$$
\Omega=\int_{- \infty}^{\infty}dx_1...\int_{- \infty}^{\infty}dx_{N-1}\int_{- \infty}^{\infty}dte^{i\sum_{n=1}^{N-1} (x_n^2-\rho^2)t}.
\eqno{(A.2)}
$$
Using
$$
\int_{- \infty}^{\infty}dxe^{ix^2t}=(-it)^{1/2} \pi^{1/2},
\eqno{(A.3)}
$$
and
$$
\lim_{\epsilon \rightarrow 0}\int_{- \infty}^{\infty}dt \frac{e^{-ixt}}{
(\epsilon-it)^{\nu}}=\left\{ \begin{array}{ll}
                           \frac{2 \pi x^{\nu-1}}{\Gamma(\nu)} & \mbox{if $x>0$,}
\\
                           0  & \mbox{otherwise,}
                           \end{array}
                            \right.
\eqno{(A.4)}
$$
we have from Eqs. (A.1) and (A.2)
$$
\Omega=2\pi \rho^{N-3}\frac{\Gamma(1/2)^{N-1}}{\Gamma((N-1)/2)},
\eqno{(A.5)}
$$
and 
$$
V(\rho)=\frac{N(N-1)}{2}\frac{\Gamma((N-1)/2)}{\Gamma(N/2-1)\Gamma(1/2)}
\frac{1}{\rho}\int_{-\rho}^{\rho}V_{int}(\sqrt{2}x)(1-\frac{x^2}{\rho^2})^{N/2-2}dx
\eqno{(A.6)}
$$
For the contact interaction $V_{int}(\sqrt{2}x)=c\delta(\sqrt{2}x)$ we have 
$$
V(\rho)=cN(N-1)\frac{\Gamma((N-1)/2)}{2\sqrt{2\pi}\Gamma(N/2-1)}\frac{1}{\rho}
\eqno{(A.7)}
$$

\pagebreak

{\bf Appendix B}

\vspace{8pt}
In this Appendix, we present an outline for evaluating the effective potential $V(\rho)$
for the isotropic case. 

Our starting formula is
$$
V(\rho)=\frac{N(N-1)}{2}\int d \vec{r}V_{int}(r) d\vec{R} d \vec{r}_3...d\vec{r}_N 
\int_{-\infty}^{\infty}dt 
e^{i(r^2/2+R^2/2)t}
e^{i(\sum_{n=3}^{N} r_n^2-\rho^2)t}/ \Omega,
\eqno{(B.1)}
$$
where
$$
\Omega=\int d \vec{r} d\vec{R} d \vec{r}_3...d\vec{r}_N  \int_{-\infty}^{\infty}
dt
e^{i(r^2/2+R^2/2)t}
e^
{i(\sum_{n=3}^{N} r_n^2-\rho^2)t}.
\eqno{(B.2)}
$$
Using
$$
\int_{0}^{\infty}x^{\nu-1} \exp(-\mu x^p)dx=\frac{1}{\mid p \mid} \mu^{-\nu/p}
\Gamma(\nu/p),
\eqno{(B.3)}
$$
and Eq. (A.4) we have from (B.1) and (B.2)
$$
\Omega=\frac{(2 \pi \Gamma(3/2))^N 16 \pi \rho^{3N-2}}{\Gamma(3N/2)},
\eqno{(B.4)}
$$
and
$$
V(\rho )=\frac{N(N-1)}{\sqrt{2 \pi}}\frac{\Gamma(3N/2)}{\Gamma(3N/2-3/2)}\frac{1}
{\rho^3}
\int_{0}^{\sqrt{2} \rho}r^2 dr (1-\frac{r^2}{2 \rho^2})^{3N/2-5/2}V_{int}(r).
\eqno{(B.5)}
$$
Now we calculate the effective potential $V(\rho)$ for various potentials 
$V_{int}$.
Substitution of
$$
V_{int}(r)=\lambda \delta(\vec{r})=\frac{\lambda}{4 \pi r^2} \delta(r)
\eqno{(B.6)}
$$
into Eq. (B.5) gives
$$
V(\rho )=\frac{N(N-1)}{2 (2 \pi)^{3/2}}\frac{\Gamma(3N/2)}{\Gamma(3N/2-3/2)}
\frac{\lambda}{\rho^3}.
\eqno{(B.7)}
$$ 
For the case of a square-well potential
$$
V_{int}(r)=\left\{ \begin{array}{ll}
              V_{0}  & \mbox{if $r \leq b$,
}
               \\
               0  & \mbox{otherwise,}
\end{array}
\right.
\eqno{(B.8)}
$$
the calculation gives the following result
$$
V(\rho )=\left\{ \begin{array}{ll}
           V_{0}/2 & \mbox{if $\rho \leq b/\sqrt{2}$,
}
 \\
 f(\rho)
  & \mbox{otherwise,}
\end{array}
\right.
\eqno{(B.9)}
$$
where 
 $$
f(\rho)= b^3 V_{0} \frac{N(N-1)}{3 \sqrt{2 \pi}} \frac{\Gamma(3N/2)}
{\Gamma(3N/2-3/2)}
\; _2F_1(3/2, ( 5-3N)/2; 5/2; b^2/(2 \rho^2))/\rho^3,
$$
 and $_pF_q$ is the 
generalized hypergeometric function.

For the Coulomb potential $V_{int}(r)=\alpha/r$ we obtain
$$
V(\rho)=\frac{2 N \Gamma(3N/2)}{3 \sqrt{2 \pi} \Gamma(3N/2-3/2)} \frac{\alpha}
{\rho}.
\eqno{(B.10)}
$$

For a Gaussian potential $V_{int}(r)=V_0e^{-\beta^2 r^2}$
$$
V(\rho)=\frac{V_0 N(N-1)}{2}\Phi(3/2,3N/2;-2\beta^2 \rho^2),
\eqno{(B.11)}
$$
where $\Phi(a,b,;x)$ is the confluent hypergeometric function.
For $2\beta^2 \rho^2 \gg 1$ Eq. (B.11) gives
$$
V(\rho) \approx \frac{V_0 N(N-1)}{2} \frac{\Gamma(3N/2)}{\Gamma(3N/2-3/2)}
\frac{1}{(\beta \rho)^3},
\eqno{(B.12)}
$$
and for $2\beta^2 \rho^2 \ll 1$ 
$$
V(\rho) \approx \frac{V_0 N(N-1)}{2}
\eqno{(B.13)}
$$

For the Yukawa potential $V_{int}(r)=V_0e^{-\mu r}/r$
$$
\everymath={\displaystyle}
\begin{array}{rcl}

&~& V(\rho)=\frac{N(N-1) \Gamma(3N/2) V_0}{\sqrt{2 \pi} \Gamma(3N/2-3/2) \rho^3}
[\frac{4\rho^2}{6(N-1)} \; _1F_2(1;1/2,3N/2-1/2;\mu^2 \rho^2/2)
\\
&~& \\
&~&
-2^{3N/4}
\mu^{2-3N/2}3(N-1)\sqrt{\pi}\rho^{4-3N/2}I(3N/2-1,\sqrt{2} \mu \rho)
\Gamma(3N/2-3/2)],
\end{array}
\eqno{(B.14)}
$$
In the $N=3$ case Eq.(B.14) simplifies
 $$
\everymath={\displaystyle}
\begin{array}{rcl}
&~& 
V(\rho)=\frac{N(N-1) \Gamma(3N/2) V_0}{\sqrt{2 \pi} \Gamma(3N/2-3/2) \mu^6 \rho^7}[(\mu \rho)^4-6(\mu \rho)^2+30
\\
&~& \\
&~&
-2e^{-\sqrt{2}\mu \rho}(
2\sqrt{2}(\mu \rho)^3+12(\mu \rho)^2+15\sqrt{2}\mu \rho+15)].
\end{array}
\eqno{(B.15)}
$$

\vspace{15 mm}
{\bf Appendix C}

\vspace{18pt}
In this Appendix, we present an outline for obtaining the effective anisotropic potential.
For an anisotropic case we introduce
$$
x=\sqrt{\sum_{n=1}^{N}x_{n}^2},\; y=\sqrt{\sum_{n=1}^{N}y_{n}^2},\; 
z=\sqrt{\sum_{n=1}^{N}z_{n}^2},
\eqno{(C.1)}
$$
we can then write for effective potential $V(x,y,z)$ the following expression
$$
\everymath={\displaystyle}
\begin{array}{rcl}
&~& V(x,y,z)=\frac{N(N-1)}{2} \int_{-\infty}^{\infty}dx^{\prime} 
\int_{-\infty}^{\infty}dy^{\prime} \int_{-\infty}^{\infty}dz^{\prime} V_{int}
(\sqrt{(x^{\prime})^2+(y^{\prime})^2+(z^{\prime})^2})
 \\
&~& \\
&~& \times \int_{-\infty}^{\infty}dR_x \int_{-\infty}^{\infty}dR_y 
\int_{-\infty}^{\infty}dR_z
\int_{-\infty}^{\infty}dx_3...\int_{-\infty}^{\infty}dx_N
\int_{-\infty}^{\infty}dy_3...\int_{-\infty}^{\infty}dy_N
 \\
&~& \\
&~& \times \int_{-\infty}^{\infty}dz_3...\int_{-\infty}^{\infty}dz_N
\int_{-\infty}^{\infty}dt_1 \int_{-\infty}^{\infty}dt_2 
\int_{-\infty}^{\infty}dt_3 e^{i[(x^{\prime})^2/2+R_x^2/2+\sum_{n=3}^{N}x_n^2-
x^2]t_1}
 \\
&~& \\
&~& \times 
e^{i[(y^{\prime})^2/2+R_y^2/2+\sum_{n=3}^{N}y_n^2-y^2]t_2}
e^{i[(z^{\prime})^2/2+R_z^2/2+\sum_{n=3}^{N}z_n^2-z^2]t_3}
\end{array}
\eqno{(C.2)}
$$
where
$$
\everymath={\displaystyle}
\begin{array}{rcl}
&~& \Omega=\int_{-\infty}^{\infty}dx^{\prime}
\int_{-\infty}^{\infty}dy^{\prime} \int_{-\infty}^{\infty}dz^{\prime}
\int_{-\infty}^{\infty}dR_x \int_{-\infty}^{\infty}dR_y
\int_{-\infty}^{\infty}dR_z
 \\
&~& \\
&~& \times \int_{-\infty}^{\infty}dx_3...\int_{-\infty}^{\infty}dx_N

\int_{-\infty}^{\infty}dy_3...\int_{-\infty}^{\infty}dy_N
\int_{-\infty}^{\infty}dz_3...\int_{-\infty}^{\infty}dz_N
\int_{-\infty}^{\infty}dt_1 
 \\
&~& \\
&~& \times
\int_{-\infty}^{\infty}dt_2
\int_{-\infty}^{\infty}dt_3
 e^{i[(x^{\prime})^2/2+R_x^2/2+\sum_{n=3}^{N}x_n^2-
x^2]t_1}
e^{i[(y^{\prime})^2/2+R_y^2/2+\sum_{n=3}^{N}y_n^2-y^2]t_2}
 \\
&~& \\
&~& \times
e^{i[(z^{\prime})^2/2+R_z^2/2+\sum_{n=3}^{N}z_n^2-z^2]t_3}.
\end{array}
\eqno{(C.3)}
$$
Using (A.4) and (B.3) we obtain
$$
\Omega=\Gamma(1/2)^{3N} (4 \pi)^3 x^{N-2}y^{N-2}z^{N-2}\Gamma(N/2)^{-3},
\eqno{(C.4)}
$$
and
$$
\everymath={\displaystyle}
\begin{array}{rcl}
&~& V(x,y,z)=\frac{N(N-1)}{2(2 \pi)^{3/2}}[\frac{\Gamma(N/2)}{\Gamma(N/2-1/2)}]^3
\frac{1}{xyz}
 \int_{-\sqrt{2}x}^{\sqrt{2}x} \int_{-\sqrt{2}y}^{\sqrt{2}y}
\int_{-\sqrt{2}z}^{\sqrt{2}z}dx^{\prime}dy^{\prime} dz{\prime}
 \\
&~& \\
&~& \times V_{int}(\sqrt{(x^{\prime})^2+(y^{\prime})^2
+(z^{\prime})^2})
[(1-\frac{(x^{\prime})^2}{2 x^2})(1-\frac{(y^{\prime})^2}{2 
y^2})(1-\frac{(z^{\prime})^2}{2 z^2})]^{N/2-3/2}
\end{array}
\eqno{(C.5)}
$$

Finally, we quote several examples of $V(x,y,z)$ for various potentials 
$V_{int}$.

For delta-potential, $V_{int}(\vec{r})=\lambda \delta(\vec{r})$ we obtain
$$
V(x,y,z)=\frac{N(N-1)}{2(2 \pi)^{3/2}}[\frac{\Gamma(N/2)}{\Gamma(N/2-1/2)}]^3
\frac{\lambda}{xyz},
\eqno{(C.6)}
$$
and for the Gaussian potential,  $V_{int}(r)=V_0e^{-\beta^2 r^2}$
we have
$$
\everymath={\displaystyle}
\begin{array}{rcl}
&~& 
V(x,y,z)=\frac{V_0N(N-1)}{2}\Phi(1/2,N/2;-2\beta^2x^2)
\Phi(1/2,N/2;-2\beta^2y^2)
 \\
&~& \\

&~& \times \Phi(1/2,N/2;-2\beta^2z^2)
\end{array}
\eqno{(C.7)}
$$

\pagebreak

\begin{center}
{\bf References}
\end{center}

\vspace{8pt}

\noindent
1. H. Feshbach and S.I. Rubinov, {\it Phys. Rev.} {\bf 98} (1955), 188.

\noindent
2. P. Appel and J. Kample, ``Fonctions  Hyprgeometriques et Hyperspheriques,"
 Gauthier-Villars, Paris, 1926.

\noindent
3. A. M. Badalyan, F. Calogero, Yu. A. Simonov, {\it Nuovo Cimento}
{\bf 68}A (1970), 572;
F. Calogero, Yu. A. Simonov, {\it Nuovo Cimento}
{\bf 67}A (1970), 641; F. Calogero et al, { \it Nuovo Cimento} {\bf
14}A (1973), 445; F. Calogero, Yu. A. Simonov, {\it Phys. Rev.} {\bf
169} (1967), 789; A. I. Baz' et al., {\it Fiz. Elem. Chastits At.
Yadra} {\bf 3} (1972), 275 [{\it Sov. J. Part. Nucl.} {\bf 3} (1972), 137
].

\noindent
4. F. Calogero, Yu. A. Simonov, and E. L. Surkov, {\it Nuovo Cimento}
{\bf 1}A (1971), 739.

\noindent
5. J. L. Bohn, B. D. Esry, and C. H. Greene, {\it Phys. Rev}. A{\bf
58} (1998), 584.

\noindent
6. P. M. Morse and H. Feshbach, ``Methods of Theoretical Physics," McGraw-Hill,
New York, 1953.

\noindent
7. Yu. A. Simonov, {\it in} ``The Nuclear Many-Body Problem" (F.Calogero and C. Ciofi degli Atti, Eds.), Bologna, 1973.

\noindent
8. M. Fabre de la Ripelle, and J Navarro, { \it Ann. Phys}. (NY) {\bf123} 
(1979), 185.

\noindent
9. J. B. McGuire, {\it J. Math. Phys}. {\bf 5} (1964), 622);
{\it J. Math. Phys}. {\bf 7} (1966), 123.

\noindent
10. C. N. Yang, {\it Phys. Rev. Lett}. {\bf 19} (1967), 1312;
{\it Phys. Rev}. {\bf 168} (1967), 1920.

\noindent
11. L. R. Dodd, {\it J. Math. Phys}. {\bf 11} (1970), 207;
               {\it Phys. Rev}. D{\bf 3} (1971), 2536;
                {\it Aust. J. Phys}. {\bf 25} (1972), 507.

\noindent
12. L.D. Faddeev, ``Mathematical Aspects of the Three-Body
Problem in the Quantum Scattering Theory," Daniel Davey and
Company, Inc., New York, 1965.

\noindent
13. A. L. Zubarev and V. B. Mandelzweig, { \it Phys. Rev}. C{\bf 52} (1995),
 509.

\noindent
14. E.P. Harper, Y.E. Kim, and A. Tubis, { \it Phys. Rev. Lett}. {\bf
23} (1972), 1533; { \it Phys. Rev}. C{\bf 2} (1970), 877; { \it Phys. Rev.}
C{\bf 2} (1970), 2455(E); { \it Phys. Rev}. C{\bf 6} (1972), 126.

\noindent
15. R. A. Rice and Y.E. Kim, {\it Few Body Systems} {\bf14} (1993), 127.

\noindent
16. E. Fermi, { \it Riverica Sci}. {\bf7} (1936), 13.

\noindent
17. N.Bogolubov, { \it J. Phys}. {\bf 31} (1947), 23.

\noindent
18. G. Baym and C. J. Pethick, { \it Phys. Rev. Lett}. {\bf 76} (1996), 6.

\noindent
19. L. Ginzburg, and L. P. Pitaevskii, { \it Zh. Eksp. Teor. Fiz}.
{\bf 34} (1958), 1240
[{\it Sov. Phys. JETP} {\bf 7} (1958), 858]; E. P. Gross, {\it  J. Math.
Phys}. {\bf 4} (1963), 195.

\noindent
20. A. Chatterjee, {\it Phys. Rep}. {\bf 186} (1990), 249.

\noindent
21. T. D. Lee, K. Huang, and C. N. Yang, {\it  Phys. Rev}. {\bf 106} (1957),
1135.

\noindent
22.  E. R. I. Abraham, W. I. McAlexander, C. A. Sackett, and R.
G. Hulet,
{ \it Phys. Rev. Lett}. {\bf 74} (1995), 1315.

\noindent
23. Y. Kagan, G. V. Shlyapnikov, and J. T. M. Walraven, {\it  Phys.
Rev. Lett}.
{\bf 76}, (1996) 2670. 

\noindent
24.  Alexander L. Fetter, cond-mat/9510037.

\noindent
25. M. Houbiers and H.T.C. Stoof, {\it Phys. Rev}. A {\bf 54} (1996), 5055.

\noindent
26. E. V. Shuryak, {\it Phys. Rev}. A {\bf 54} (1996), 3151.

\noindent
27. M. Ueda, and A. J. Leggett, {\it Phys. Rev. Lett}. {\bf 80} (1998), 1576.

\noindent
28.  F. Dalfovo and S.Stringari, {\it Phys. Rev}. {\bf A53} (1996),  2477.

\noindent
29.  R. J. Dodd, M. Edwards, C. J. Williams, C. W. Clark, M. J.
Holland,
P. A. Ruprecht, and K. Burnett, {\it Phys. Rev}. {\bf A54} (1996), 661.

\noindent
30. Y. E. Kim and A. L. Zubarev, {\it Phys. Lett}. A{\bf 246} (1998), 389.

\noindent
31. M. Wadati and T. Tsurumi, {\it Phys. Lett}. A{\bf 247} (1998), 287.

\noindent
32.   C. C. Bradley, C. A. Sackett, and R. G. Hulet, {\it Phys. Rev.
Lett}.
{\bf 78} (1997), 985.

\end{document}